\documentclass[runningheads]{llncs}
\usepackage[T1]{fontenc}
\usepackage{graphicx}
\usepackage{subcaption}
\usepackage{hyperref}
\usepackage{xurl}

\usepackage[margin=1in]{geometry}
\usepackage[utf8]{inputenc}
\usepackage{lmodern}
\usepackage{setspace}
\usepackage{booktabs}
\usepackage{longtable}
\usepackage{array}
\usepackage{float}
\usepackage{amsmath}
\usepackage{natbib}
\date{}

\begin{document}
\begin{sloppypar}
\title{Interpupillary Distance Constraints in Pediatric VR: Implications for Psychology and Psychotherapy}

%
%\titlerunning{Abbreviated paper title}
% If the paper title is too long for the running head, you can set
% an abbreviated paper title here
%
\author{
Grzegorz Pochwatko\inst{1}\orcidID{0000-0001-8548-6916}
}
\authorrunning{G. Pochwatko}
% First names are abbreviated in the running head.
% If there are more than two authors, 'et al.' is used.
%
\titlerunning{Interpupillary Distance - Psychology, Psychotherapy \& Pediatric VR}

\institute{
Virtual Reality and Psychophysiology Lab\\
Institute of Psychology, Polish Academy of Sciences\\
Warsaw, Poland}
\maketitle              % typeset the header of the contribution
\begin{abstract}
Virtual reality (VR) is increasingly used across psychology, from experimental research and assessment to counseling, psychological treatment, and psychotherapy, with growing applications for children and adolescents in areas such as exposure-based intervention, emotion regulation training, pain and anxiety reduction, rehabilitation, psychoeducation, and studies of perception, attention, and embodiment. In these contexts, VR is often treated as a relatively neutral delivery medium: a tool assumed to present the same procedure to all participants. This assumption may be misleading. Most consumer head-mounted displays (HMDs) have been designed primarily for adult anthropometry, including adult interpupillary distance (IPD) ranges. As a result, some children may be excluded from participation, or may receive a systematically degraded perceptual experience, not because of psychological ineligibility, but because the device cannot be adequately aligned to their visual anatomy and may further present age-sensitive ergonomic limitations related to headset size, facial fit, and weight.

This paper argues that IPD constraints in consumer VR headsets represent an underrecognized methodological and clinical problem in pediatric psychology and psychotherapy. More broadly, this concern extends beyond optical alignment alone to age-sensitive ergonomic factors such as headset size, facial fit, and weight. If headset fit affects binocular fusion, visual comfort, depth perception, attentional load, cybersickness, willingness to remain in the simulation, and subjective sense of presence, then it may also influence core psychological variables such as engagement, emotional processing, distress tolerance, dropout, and treatment response \citep{Hibbard2020,MonWilliams1993,Eijlers2019,Shahid2024}. In this way, the headset may function as a selection mechanism, shaping who is included in studies, who can tolerate intervention, and to whom findings can legitimately be generalized. Such selection may at times be appropriate on methodological or clinical grounds, but it should be made explicit rather than remaining an unmeasured and easily misinterpreted feature of recruitment and intervention delivery.

Using published developmental data on IPD, official headset specifications, and examples from pediatric and youth-oriented VR studies, we show that anthropometric mismatch is likely to disproportionately affect younger children and those at the lower end of the IPD distribution \citep{MacLachlan2002,MetaQuest3Official,MetaIPDHelp,Kim2021IPD,Zhao2023Contrast,Tamura2024}. We use Meta Quest 3 as a contemporary case study because it is among the more accommodating widely available consumer HMDs, yet still appears likely to leave a meaningful portion of younger children near or below the margin of comfortable use. We propose that pediatric VR research and therapy should treat headset compatibility not as background technical detail, but as part of psychological method. At minimum, studies should report headset model, optical adjustment range, IPD measurement procedures, and exclusions or attrition plausibly related to headset mismatch. More broadly, the paper calls for a psychologically informed framework for evaluating whether VR-based procedures are truly comparable, accessible, and generalizable across developmental groups.
\end{abstract}

\section{Introduction}

Virtual reality is becoming an important tool in psychology and psychotherapy. It is used not only for experimental work, but also for applied and clinical purposes, including pain distraction, anxiety reduction, rehabilitation, skills training, psychoeducation, and exposure-based interventions \citep{Won2017,Wang2021,KimKim2020,Olbrecht2021}. In child and adolescent populations, VR is especially attractive because it offers standardization, controllability, repeatability, and high engagement. For these reasons, VR is often presented as a promising platform for delivering psychological procedures in a controlled yet ecologically vivid way.

This promise rests on an implicit assumption: that the technology delivers roughly the same intervention to all participants who are nominally eligible. In practice, this may not be true. Consumer VR headsets are not psychologically neutral containers. They are physical devices with optical constraints, weight, fit requirements, and anthropometric assumptions. When such devices are used with children, these assumptions matter because children’s bodies, faces, and visual parameters differ systematically from those of adults. One of the most important of these parameters is interpupillary distance (IPD), the distance between the centers of the pupils.

This paper advances a simple but consequential claim: consumer VR headsets may introduce a hidden source of bias into pediatric psychological research and VR-assisted psychotherapy. If a child’s IPD falls outside the optical range of the device, the child may be excluded altogether. If the child falls near the lower edge of the range, the intervention may still be delivered under less comfortable or less stable perceptual conditions. In either case, the resulting study or therapeutic protocol may no longer apply equally to the whole target population. This may alter engagement, sense of presence, visual comfort, attentional allocation, emotional processing, and adherence \citep{Hibbard2020,Shahid2024,Eijlers2019}. It therefore has methodological and clinical implications.

The central concern of this paper is therefore broader than headset fit alone. The question is whether VR systems used in developmental and clinical psychology are introducing an unmeasured selection mechanism into research and practice. If so, then some current findings in pediatric VR may be less generalizable than they appear, and some clinical applications may be less universally accessible than assumed.

\section{Why IPD matters psychologically, not only optically}

Interpupillary distance may appear to be a narrow technical parameter of headset configuration. In psychological and psychotherapeutic applications of VR, however, its implications are broader. When the optical system is poorly aligned to the user’s IPD, the result may include off-center viewing, reduced binocular comfort, less stable stereoscopic fusion, visual strain, and altered depth perception \citep{Hibbard2020,MonWilliams1993,Lambooij2009}. These effects influence perceptual quality, but they may also propagate into variables that psychologists actually care about by introducing systematic error into measurement and intervention delivery.

In \textbf{research settings}, this may affect attention, task performance, fatigue, tolerance for longer sessions, and subjective ratings of presence or immersion. In \textbf{clinical settings}, the consequences may be even more important. A child who sees the scene less comfortably may disengage faster, feel overwhelmed sooner, report stronger nonspecific discomfort, or attribute strain to the intervention itself. In \textbf{exposure-based interventions}, one must distinguish therapeutic activation from discomfort generated by the medium. In \textbf{psychoeducation, mindfulness, or regulation protocols}, headset-induced strain may compete with the intended psychological task. Thus, an IPD mismatch can act as a confound: what appears to be a difference in distress tolerance, susceptibility to cybersickness, treatment adherence, or acceptability may partly reflect differential hardware fit \citep{PriceAnderson2007,Hur2021,Shahid2024}.

For this reason, the present paper treats IPD constraints as a psychological methodology issue rather than as a narrow engineering concern. In pediatric VR, hardware fit may shape not only whether a child can participate, but what kind of psychological experience the child is actually having.

\section{Developmental change in IPD and pediatric eligibility}

Interpupillary distance changes substantially across development. Classical work by MacLachlan and Howland \citet{MacLachlan2002} reported age-related values and standard deviations for children and adolescents from 1 month to 19 years, showing systematic growth of IPD across childhood and adolescence. This matters because pediatric samples are not simply smaller versions of adult samples. They are developmentally heterogeneous, and the lower tail of the IPD distribution is concentrated in younger children.

In practice, this means that the compatibility of a given headset with a pediatric population is not binary. It is probabilistic and age-sensitive. A headset whose lower IPD boundary accommodates most adolescents may still be poorly suited to younger school-age children. Even within a nominally eligible age range, some children may be comfortably aligned while others are fitted at the margin of the optical sweet spot or outside it altogether.

This age dependence is especially relevant in psychological work because the intended target populations are often defined developmentally rather than anthropometrically. Researchers may recruit children with autism spectrum disorder, children with ADHD, or preadolescents with anxiety, while the actual deliverability of the intervention may still depend partly on unmeasured optical fit. As a result, developmental eligibility and hardware eligibility should not be treated as equivalent.

\section{How headset constraints may bias psychological samples}

The sampling problem emerges when anthropometric compatibility quietly shapes who enters, completes, or benefits from a study. Some children may be excluded before enrolment if their IPD lies outside the adjustable range of the headset. Others may pass initial screening but experience poorer comfort, more strain, or reduced engagement during the task. These participants may withdraw, provide lower-quality data, or show reduced adherence to repeated sessions.

Methodologically, the final sample may not represent the intended population, but only the subset compatible with the device. In developmental research, this can bias age comparisons by selectively retaining younger participants with relatively large IPD values for their age. In clinical studies, it can make an intervention appear more tolerable or more effective than it would be in the broader target group. In both cases, hardware compatibility becomes part of sampling and should be treated as part of the method rather than as incidental device metadata.

\section{Quest 3 as a case study in pediatric psychological accessibility}

Meta Quest 3 is a useful case study because it is widely available, relatively advanced, and more accommodating than several earlier consumer headsets. According to official manufacturer information, Quest 3 mechanically accommodates interpupillary distances from 53 to 75 mm \citep{MetaQuest3Official}. Meta also states that its headsets work best for users with IPDs between 56 and 70 mm \citep{MetaIPDHelp}. The two ranges suggest that mechanical accommodation and comfortable operation are not identical.

A child whose IPD is just above the lower mechanical limit may be technically able to use the headset while still remaining near the lower edge of comfortable alignment. Accordingly, ``usable'' should not be assumed to mean equivalent across participants.

Quest 3 is a useful contemporary example: more accommodating than many earlier consumer headsets, but still likely to leave a nontrivial proportion of younger children near or below the edge of comfortable fit. Even with the lower limit of 53 mm, the updated estimates suggest that compatibility remains incomplete in younger age groups, with approximately 44\% of 6-year-olds, 56\% of 7-year-olds, and 71\% of 8-year-olds expected to fall within range.

\section{Illustrative compatibility estimates across childhood}

The estimates reported here were derived from the sex-specific corrected PD means, standard deviations, and sample sizes reported in MacLachlan and Howland (2002). For each age group, the estimated mean IPD was calculated as the sample-size-weighted mean of the female and male values, whereas the estimated 10th percentile and the estimated proportion of children with IPD $\geq 53$ mm were derived from an age-specific two-component normal-mixture model.

\begin{table}[H]
\centering
\caption{Illustrative compatibility of Quest 3 with pediatric IPD distribution.}
\label{tab:compatibility}
\begin{tabular}{cccc}
\toprule
Age (years) & Estimated mean IPD (mm) & Estimated 10th percentile (mm) & Estimated \% with IPD $\geq$ 53 mm \\
\midrule
6  & 52.6 & 49.0 & 44.1 \\
7  & 53.4 & 49.6 & 55.8 \\
8  & 54.6 & 50.9 & 71.0 \\
9  & 54.9 & 51.6 & 76.5 \\
10 & 56.2 & 52.4 & 86.0 \\
11 & 57.1 & 53.5 & 92.8 \\
12 & 57.6 & 53.3 & 91.6 \\
13 & 58.3 & 53.5 & 91.9 \\
14 & 59.8 & 55.0 & 96.6 \\
\bottomrule
\end{tabular}
\vspace{0.5em}
\parbox{0.95\linewidth}{\footnotesize \textit{Note.} Estimated values were derived from the sex-specific corrected PD means, SDs, and sample sizes reported by \citet{MacLachlan2002}.}
\end{table}

\begin{figure}[H]
    \centering
    \includegraphics[width=0.95\textwidth]{quest3_pediatric_ipd_compatibility_updated_from_maclachlan2002.png}
    \caption{Estimated Quest 3 compatibility across childhood. The main curve shows the estimated proportion of children with IPD $\geq$ 53 mm. Shaded zones indicate a heuristic interpretation of practical suitability. Secondary lines show illustrative anthropometric and ergonomic context.}
    \label{fig:quest3_compatibility}
\end{figure}

Table~\ref{tab:compatibility} and Figure~\ref{fig:quest3_compatibility} suggest that, particularly in younger cohorts, the subgroup accessible to Quest 3 may be narrower than the nominal target population. Unless headset fit is measured and reported, this affects both recruitment and interpretation.

\section{The problem in pediatric and youth VR studies}

Pediatric and youth-oriented VR studies already identify feasibility, safety, comfort, and tolerability as determinants of participation and outcome \citep{Wang2021,Tennant2020,Tennant2021,Verheul2022}. Even when IPD is not explicitly reported, hardware usability and adverse effects remain part of the clinical and methodological context.

More broadly, the pediatric VR literature consistently frames VR as a psychologically meaningful tool for pain reduction, procedural anxiety, exposure, psychosocial support, and rehabilitation \citep{Won2017,Eijlers2019,Wong2020,Schmitt2011,Olbrecht2021}. This gives fit-related exclusion direct methodological relevance. If the medium itself is selectively less usable for younger or smaller children, then observed feasibility and efficacy may partly reflect hardware compatibility rather than psychological suitability.

A second sign comes from the wider VR perception literature. Studies on small-IPD users and headset image quality show that adult-centered assumptions do not generalize. IPD mismatch can degrade image quality, affect contrast perception, and alter depth-related judgments, with special relevance for users at the lower end of the IPD range \citep{Kim2021IPD,Zhao2023Contrast,Bhansali2024,Tamura2024}. These effects provide a plausible mechanism by which a child may remain formally eligible for a VR study while nevertheless receiving a meaningfully different perceptual experience.

\section{Implications for research validity in psychology}

The first implication is for internal validity. If poorer headset fit increases nonspecific discomfort, attentional distraction, or difficulty sustaining presence, then outcomes attributed to a psychological manipulation may be partly shaped by the quality of optical alignment. Presence, fear activation, habituation, distress tolerance, or engagement may not be fully separable from hardware fit \citep{PriceAnderson2007,Hur2021,Morina2014}.

The second implication is for external validity. Findings from samples filtered by headset compatibility may not generalize to all children in the intended age or clinical range. This is especially problematic in developmental comparisons and in feasibility studies, where conclusions about acceptability or tolerability are often made from relatively small samples.

The third implication concerns theory. If VR is used to study psychological processes such as embodiment, spatial cognition or emotional regulation under controlled conditions, then systematic fit-related variation may enter the measurement model. In other words, the hardware can influence not only who is included, but what psychological construct is actually being measured.

\section{Implications for psychological intervention and psychotherapy}

The consequences of headset mismatch are especially important when VR is used as part of psychological intervention. In psychotherapy and related clinical applications, VR is valued because it can standardize emotionally relevant situations, regulate the intensity of exposure, and support repeated practice under controlled conditions \citep{KimKim2020,Chesham2018,Beele2024}. These advantages assume that the medium itself is sufficiently stable across users. If some children receive a substantially less comfortable or less immersive experience, then comparability across participants is weakened at the level of the therapeutic process.

For instance, in exposure-based work, discomfort arising from optical mismatch may be misread as anxiety triggered by the feared stimulus, making interpretation of subjective distress more difficult. In regulation-focused interventions, such as breathing, mindfulness, grounding, or attentional training in VR, headset-related strain may undermine the very skills being practiced \citep{Olbrecht2021,Chuan2023}. In cognitive or psychoeducational protocols, fatigue and reduced comfort may impair concentration and compliance. In all these cases, the relevant problem is not only whether the child can technically wear the headset, but whether the child is receiving the psychological intervention in a functionally equivalent way.

For that reason, anthropometric compatibility should be treated as part of treatment accessibility. Just as clinicians consider language level, sensory sensitivity, developmental stage, and family context, they may also need to consider whether the visual and ergonomic properties of the headset allow the child to engage with the intervention as intended. If not, then VR-assisted psychotherapy risks becoming selectively accessible to children whose bodies fit the device rather than to the broader clinical population for whom the intervention is theoretically intended.

\section{Cybersickness, tolerability, and the interpretation of pediatric outcomes}

Cybersickness and visual fatigue are relevant to the same methodological problem. Pediatric VR studies generally report favorable tolerability when sessions are carefully designed and clinically supervised \citep{Tennant2020,Tennant2021,Stunden2020}. However, tolerability is itself partly shaped by the usability and fit of the device. A child who experiences visual strain, dizziness, or unstable viewing may be less likely to complete the task or benefit from the intervention, even if the nominal clinical content is well designed.

Studies of pediatric VR often interpret low distress, low dropout, or high satisfaction as evidence of feasibility. Such conclusions are valuable, but they may be sample-dependent if children most vulnerable to poor fit are underrepresented from the outset. In this sense, optical compatibility and tolerability are not separate issues; they jointly influence who remains in the sample and what conclusions are drawn about acceptability.

\section{Recommendations for reporting and clinical screening}

At minimum, pediatric VR studies should report the headset model and the stated IPD range of the device. If IPD is measured, authors should describe the measurement method, summarize the sample distribution, and report any IPD-based exclusions. If IPD is not measured, this should be acknowledged as a methodological limitation.

Studies should also report whether participants were screened out before enrolment because of headset fit, whether fitting problems emerged during exposure, and whether sessions were terminated or shortened because of discomfort, blurred fusion, eye strain, or nausea. In intervention studies, hardware-related attrition should be distinguished from treatment-related dropout whenever possible.

In clinical settings, a brief pre-use screening protocol may be warranted, including at minimum assessment of headset fit, perceived comfort, whether the child can maintain clear and stable viewing without excessive strain, and, where relevant, susceptibility to cybersickness \citep{OsekaPochwatkoSwidrak2023}. Such screening does not solve the broader population-level issue, but it can reduce the risk of misinterpreting hardware mismatch as poor motivation, poor distress tolerance, or lack of treatment suitability.

We did not identify a dedicated consensus standard for reporting IPD-related exclusions in pediatric VR studies, so these recommendations should be understood as an evidence-informed methodological proposal built from the convergence of developmental IPD norms, VR optics research, and pediatric clinical VR feasibility work \citep{MacLachlan2002,Hibbard2020,Eijlers2019,Wang2021,Tennant2021}.

\section{Recommendations for device selection and future development}

For researchers and clinicians selecting hardware, lower minimum IPD should be treated as a meaningful psychological-access variable, not merely a technical specification. A headset that accommodates a broader range of children may increase representativeness and reduce hidden exclusion. However, broader nominal range alone is not enough. The literature on image quality, contrast, optical aberrations, eye box design, and stereoscopic perception suggests that pediatric usability depends on a broader set of optical and ergonomic parameters than IPD alone \citep{Bhansali2024,Zhao2023Contrast,Zhao2024Quality,Le2023EyeBox,Tamura2024}.

More broadly, the field may need to move beyond the assumption that adult-oriented consumer headsets are automatically adequate for pediatric psychological work. If VR is to become a serious tool in developmental psychology and psychotherapy, there is a strong argument for pediatric-informed design standards and for more explicit reporting of anthropometric compatibility in both research and practice.

\section{Limitations}

This paper is a methodological and conceptual analysis rather than a direct headset-fitting study in a newly measured pediatric cohort. The compatibility estimates are illustrative and based on published age-group corrected PD data. Although the analysis focuses on interpupillary distance as a clear and measurable marker of compatibility, pediatric accessibility also depends on broader optical and ergonomic factors, including facial geometry, eye relief, strap configuration, headset size and weight, task duration, and whether the child remains centered in the optical sweet spot \citep{DAmato2022,Deng2011}. The argument therefore identifies one central source of hidden exclusion rather than the full range of compatibility constraints in pediatric VR.

The core claim is practical: in pediatric psychology and psychotherapy, consumer headset fit is likely to function as an undermeasured moderator of access, comfort, and generalizability. Compatibility should not be treated as all-or-none, and children above a nominal threshold should not be assumed to have equivalent experiences.

A second limitation concerns the literature itself. Direct pediatric HMD anthropometry and fit studies remain scarce. Much of the strongest direct evidence comes from developmental IPD norms, optics/perception work, and pediatric VR feasibility studies rather than from large-scale pediatric headset-fit validation studies. That gap is precisely part of the problem identified here.

\section{Conclusion}

In pediatric psychology and psychotherapy, virtual reality is often discussed as if it were a standardized therapeutic or research medium. This paper argues that such a view is incomplete. Consumer VR headsets embed anthropometric assumptions that may systematically privilege some children over others, especially when interpupillary distance falls near or below the lower optical range of the device. As a result, VR may introduce an invisible layer of selection into psychological research, assessment, and intervention.

The point is that the field should become more psychologically precise about what counts as access to the intervention. A child who is nominally enrolled in a VR study but receives a visually compromised or uncomfortable experience is not necessarily receiving the same psychological procedure as another participant. Likewise, a treatment protocol that works well only for children whose facial and visual anatomy match adult-oriented hardware cannot automatically be assumed to generalize to the whole target population.

If VR is to mature as a serious tool in developmental psychology, clinical psychology, and psychotherapy, then headset compatibility must be treated as part of method, not as a hidden background condition. Measuring and reporting IPD, documenting hardware-related exclusions, and acknowledging the possibility of hardware-mediated sampling bias are small methodological steps that could substantially improve the validity, transparency, and clinical credibility of pediatric VR research.

\bibliographystyle{apalike}
\bibliography{bibliography}

\end{sloppypar}
\end{document}